# Generalization with Reverse-Calibration of Well and Seismic Data Using Machine Learning Methods for Complex Reservoirs Predicting During Early-Stage Geological Exploration Oil Field

Dmitry Ivlev

dm.ivlev@gmail.com

**The aim of this study** is to develop and apply an autonomous approach for predicting the probability of hydrocarbon reservoirs spreading in the studied area.

**Materials and methods.** The prediction was made based on the 3D seismic survey data and well information on the early exploration stage of the studied field. The results of the lithological interpretation of logging from nine wells were used, four of which penetrated the object vertically or subvertically, while the remaining five were drilled horizontally through different stratigraphic parts of the Achimov sedimentary complex, which is the object of this study. The paper presents an approach based on a technological stack of machine learning algorithms with the task of binary classification and modification of the geological-geophysical dataset. The study includes the following sequence of actions: creation of data sets for training, selection of features, reverse-calibration of data, creation of a population of classification models, evaluation of classification quality, evaluation of the contribution of features in the prediction, ensembling the population of models by stacking method.

**As a result,** a prediction was made - a three-dimensional cube of calibrated probabilities of belonging of the studied space to the class of reservoir and its derivative in the form of the map of reservoir thicknesses of the Achimov complex of deposits was obtained. Assessment of changes in the quality of the forecast depending on the use of different data sets was carried out.

**Conclusion.** The reverse-calibration method proposed in this work uses the uncertainty of geophysical data as a hyperparameter of the global tuning of the technological stack, within the given limits of the a priori error of these data. It is shown that the method improves the quality of the forecast. The technological stack of machine learning algorithms used in this work allows expert-independent generalization of geological and geophysical data, and use this generalization to test hypotheses and create geological models based on a probabilistic view of the reservoir. The approach, formalizes the generalization of data about the target, using only factual information such as lithology along the wellbore and seismic data. Depending on the input data, the approach can be a useful tool for finding and exploring geologic targets, identifying potential resources, and optimizing and designing reservoir development systems.

## Introduction

This paper develops a methodology for predicting the spatial distribution of rocks that are reservoirs for hydrocarbons. The methodology uses machine learning algorithms in the problem of binary classification, which restore the probability function of the space element belonging to the classes identified by the results of interpretation of well logging. Attributes of seismic wavefield are used as predictors.

The peculiarities of the study include the following factors:

- the object of the study is the Achimov complex of deposits, the formation of which is associated with the flow to the foot of the Neocomian shelf terraces of sand and silt sediments in the form of turbidite flows of varying density and landslides;
- the forecast is made on the basis of 3D seismic survey data at the early stage of exploration works, when most of the study area is not covered by drilling results, and the well information about the study object is mainly concentrated in one area;
- the number of wells that penetrated the object of study is nine, of which four wells penetrated the object vertically or subvertically, and five - with horizontal penetration in different stratigraphic parts of the studied complex of deposits.

Given the features described above, the purpose of this research is to develop and demonstrate an expert-independent-autonomous technology stack that consists of machine learning algorithms and methods for modifying training data to predict the probability of reservoir propagation. Autonomy means that after preparation and input of geological and geophysical information, as well as determination of uncertainty parameters and a priori calibration limits, the expert's influence on algorithms operation is minimal. The optimal case is when the algorithm itself on the basis of constraints and input data determines model parameters, selects attributes, type and degree of modification of borehole and seismic data.

The study includes the following sequence of actions: creation of a basic data set, selection of features, creation of a modified data set by reverse-calibration, creation of a population of classification models, evaluation of forecast quality, evaluation of the contribution of features to the forecast, combining models into an ensemble, final forecast, analysis of the obtained results.

## Materials and methods

The area of work was covered by 3D seismic survey. After processing and complex interpretation of seismic data, a depth cube of amplitudes was obtained, from which the study area was identified. The width of the area is 12,000 meters, the length is 24,000 meters and the thickness is 400 meters. The resolution of the allocated seismic survey is 25 meters laterally and 5 meters vertically. The total dimensionality is i 480, j 960, k 80.

Using the seismic survey data, 456 seismic wavefield attributes were obtained by applying 25 different attribute extraction algorithms with different window sizes: 10, 20, 30 and 50. In addition, mono-frequency cubes were created using a spectral decomposition algorithm in the frequency channel range from 1 to 45 Hz with decomposition of frequencies by Ricker and Morlet wavelets. The space attributes were standardized, and a Yeo-Johnson stepwise transformation was also performed.

The results of lithologic interpretation of geophysical well surveys were classified into two classes, reservoir and non-collector. The reservoir was coded as 1 and the non-collector was coded as 0. The results of downhole data classification were approximated onto a grid using the dominant frequency of the class in each grid element (voxel) of the seismic cube intersecting the well trajectory. This created vectors from the spatial data, where each class definition along the well path was assigned 456 space features.

In order to create a training sample, the selection of traits was carried out in two stages. At the first stage, the features that were strongly correlated with other features, where the Pearson coefficient exceeded 0.95, were removed. At the second stage, the BoostARoota algorithm [1] based on the CatBoost classifier [2] was used to evaluate the significance of the features. As a result of the selection, 54 most significant attributes were selected from the initial data set containing 456 attributes, on the basis of these attributes was carried out further study of the data.

To modify the well data, the study applied the reverse-calibration method, which was used in [3] to determine the optimal position of the seismic wavefield relative to well trajectories by finding the best machine learning model feedback to recover a continuous sequence of lithology classes along the wellbore, using the surrounding space attribute values. In this study, this method was modified to account for the nature of the studied object.

Thus, as a result of exploratory data analysis, it was found that when lithology was approximated on a grid with a seismic measurement scale (25x25x5 m) for horizontal wells, the prediction was unstable and the classification quality score was low. The size of the seismic cube grid element to which the lithology class was approximated could not sufficiently correctly convey the variability of the environment for prediction purposes. Following the principle of autonomy of the prediction process, additional functions were incorporated into the reverse-calibration method to optimize the process. In addition to finding the best position of the well profile in the wavefield, functions were added to find the optimal level of grid sampling.

To implement the reverse-calibration method, the well trajectories were copied together with the classification well logs with lateral spacing of 0, 10 and 20 meters in different directions of light, as well as with vertical up/down spacing of 0, 5 and 10 meters. Thus, 125 trajectories were obtained for one well, including the initial position. For each trajectory, the lithological classification log was projected onto grids with different lateral spacing: 25 m (x1), 12.5 m (x2) and 6.25 m (x4). Thus, 750 variations were obtained for each well.

Finding the optimal seismic wavefield offset position relative to the original trajectory was accomplished by enumerating the variants created in the previous step and optimizing the classification error function of the machine learning algorithm on cross-validation for each unique combination of well trajectories.

Variant optimization of a group of wells for which a selection is made from a unique combination of variants of spatial position and type of approximation:

$$X_{d,l} = V(X_{D,L}),$$

$$d \in D, l \in L,$$

where $X_{D,L}$ is an array of all positions and lithology approximation options for wells D, with spatial data vectors L, function $V(\cdot)$ randomly or through a search grid generates $X_{d,l}$ – an array of a unique combination of position and approximation options for wells d with a feature vector array for this combination.

$$X_{d,l}^{min} = argmin_{X_{d,l}} [E(C(V(X_{D,L})^n))],$$

$$n \in N,$$

where each combination n from the number of well position combinations $N$ is given to the classification function $C(\cdot)$, the result of the classification is evaluated by the error function $E(\cdot)$. $X_{d,l}^{min}$ an array of a set of wells and their vectors with the minimum value of classification error.

Given the significant number of combinations possible as a result of discrete displacement of trajectories, and various approximation options, the enumeration of all possible combinations can take a long time. In this regard, the search optimizer - Parcene Tree Estimator (TPESampler) [4, 5] was applied. This method reduces the search time for optimal combinations by using previous results and a priori knowledge of the error function. TPESampler is based on building a Gaussian process model that predicts the error function for different combinations of parameters, and then selects the most promising options for the next search iterations. The Random Forest algorithm was used to evaluate the classification quality, which achieves a good balance of accuracy and speed when dealing with a large number of features. Classification quality was evaluated using the ROC AUC metric on cross-validation by wells, in which one well was extracted from the training set and the classifier was trained on the remaining wells. The quality metric of the trained classifier was then evaluated on the withdrawn well, after which the data from the withdrawn well were returned to the overall sample, the next well was selected from it, and the evaluation process was repeated. The results of the ROC AUC quality score were averaged across the wells for the overall evaluation of the offset well combinations. The ROC AUC estimate for the base case was 0.604, for the best found case 0.721. Seismic wavefield offset distances and well data approximation coefficients for the best response of the training model are shown in Table 1.

Table 1. Seismic field shifts relative to well trajectories and degree of lithology approximation to grid

| Well | X | Y | Z | Scl_down |
|---|---|---|---|---|
| 16G | -20 | -10 | 10 | x4 |
| 13G | -20 | 10 | 10 | x2 |
| 14G | -20 | 0 | 10 | x4 |
| 12G | 20 | 20 | 5 | x2 |
| 17G | -10 | 10 | 0 | x4 |
| 170 | 10 | 0 | 5 | x1 |
| 165 | 10 | 10 | -5 | x1 |
| 171 | 10 | 10 | 0 | x1 |
| 11 | 0 | 10 | 0 | x1 |

Two datasets were created: the Base dataset (Base) and the reverse-calibrated dataset (RC). The Base dataset contains wells in their original positions, approximated by the seismic survey scale. The RC set contains modified data on the wells after the reverse-calibration algorithm works. The number of class labels is given in Table 2.

Table 2. Quantity of datasets for training

| Class | Base | RC |
|---|---|---|
| 0 | 486 | 861 |
| 1 | 217 | 464 |
| Sum | 703 | 1325 |

To prevent data leakage - the accidental exchange of any information between test and training sets - a data handling protocol [3] was used, which has the following design. The data were first standardized, and then a Yeo-Johnson stepwise transformation was performed. The transformation parameters were transferred from the seismic field attribute cubes to the well data sets. Eight partitioning options were created for the training, validation, and test samples for the baseline and RC datasets. For each dataset, a variant of the same well combinations was generated. The test and validation portions included data from one vertical well or one horizontal well. Thus, the test well was isolated and used to assess classification quality, and the validation well was used to fine-tune the models and control training. Combinations of 8 partitioning options ensured that the test well data did not fall into the training sample in each individual option, but the entire data set in the various training combinations was available for the 8 models of the same machine learning algorithm in total, and they were tuned for the full range of geologic conditions penetrated by the wells.

The training design consisted of two phases. The first phase trained the baseline models and the second phase trained the metamodel on the results of the baseline models.

Training data are class labels with seismic field attributes, which is a typical example of structured tabular data. To classify such data in practice, gradient binning algorithms over decision trees (GBDT) in various implementations such as CatBoost [2], LightGBM [6] and XGboost [7] are used. Recently, however, deep learning algorithms that also perform well on tabular data have emerged. One of such algorithms, TabPFN (Prior-data Fitted Network) [8], was used in this study, which is well suited for ensemble with the results of GBDT algorithms, since its errors do not correlate with the errors of these methods (Rashka 2022).

For all machine learning algorithms, hyperparameters were optimized by maximizing the ROC AUC metric on crossvalidation with class ratio balancing. For GBTD models, the selection of hyperparameters was performed using the TPESampler optimizer [4, 5]. For TabPFN, by sequentially enumerating from 2 to 200 the parameter N_ensemble_configurations, for which the maximum ROC AUC value was achieved at a value of 21.

Using the GBDT model population, we assessed the importance of the features for the prediction (Table 3). The evaluation was performed by calculating the Shepley index for each batten separately [9]. The obtained results were scaled (min-max) within a batch, and then summed up for each attribute. Table 3 shows the 20 traits with the maximum total contribution. The dominance of the importance for prediction of the attributes obtained by the spectral decomposition algorithm with decomposition frequencies from 1 to 13 Hz by the Morlet wavelet is noted. It is likely that this algorithm can be used to partially rationalize the prediction results obtained from the 54 dimensional attribute space. At the same time, the predominance of low frequencies of amplitude decomposition by spectral decomposition in 20 significant attributes may indicate peculiarities of seismic data processing.

Table 3: Twenty most important features for GBDT algorithms

| Features | Score |
|---|---|
| SD-morlet-2 | 19.00926 |
| gradient_magnitude | 11.02174 |
| SD-morlet-4 | 10.57292 |
| SD-morlet-10 | 10.22294 |
| SD-morlet-13 | 9.849514 |
| sweetness | 9.758056 |
| SD-morlet-12 | 9.590384 |
| instantaneous_amplitude | 9.458229 |
| quadrature_amplitude | 9.351962 |
| amplitude_spectrum | 9.272130 |
| SD-ricker-2 | 9.107664 |
| SD-ricker-3 | 8.951618 |
| SD-ricker-4 | 8.749442 |
| SD-morlet-7 | 8.669727 |
| SD-morlet-5 | 8.238869 |
| dominant_frequency | 8.028878 |
| SD-ricker-5 | 8.000961 |
| SD-morlet-1 | 7.896709 |
| coherence | 7.442584 |
| SD-morlet-11 | 7.330567 |

For the final prediction, the model population was ensemble-trained using the stacking method. This ensemble learning method uses a metamodel that is trained and makes a prediction based on the prediction results of the underlying models. A logistic regression algorithm was used as the metamodel. Before ensemble training, the models were pre-trained on the full dataset with a fixed number of iterations with a limit of tree growth in depth.

The small amount of training data did not allow us to create a separate test sample for the metamodel without losing the quality of the final prediction, so we used the method of weighting votes to evaluate the metamodel. To do this, the weight was first calculated:

$$w_i = e^{k_i},$$

where $k_i$ is the coefficient in the logistic regression equation for the predictions of each individual base model. Then the weighted average value was calculated:

$$\bar{x} = \frac{\sum_{i=1}^{n} w_i \bar{x}_i}{\sum_{i=1}^{n} w_i},$$

where $\bar{x}_i$ is the classification quality metric of each individual base model, determined on the test set; n is the population of models.

The method of weighting of votes allows to estimate approximately the metamodel structure and to average the forecast quality through weighting of metrics of individual models in the ensemble.

According to the value of logistic regression coefficient the most significant models for prediction were determined: the models trained on the basic data set (Table 4) and after reverse-calibration (Table 5). The table for each model shows the degree of its contribution to the metamodel prediction (impotant scr), shows the algorithm by which the model was trained (algo), and the wells on which it was validated (val) and tested (test), gives the quality metrics of F1 score measure classification for each class (Class 0, Class 1). The final values (weighted impotant) are obtained as the weighted average of the model significance factor (impotant scr) for each quality metric.

Table 4: The most important models on the basic dataset for the metamodel and their characteristics

| Impotant scr | algo | test | val | Class 0 | Class 1 |
|---|---|---|---|---|---|
| 30.585 | cat | 16G | 171 | 0.583 | 0.563 |
| 19.084 | lgb | 12G | 165 | 0.583 | 0.578 |
| 15.752 | lgb | 171 | 12G | 0.605 | 0.522 |
| 13.237 | xgb | 12G | 165 | 0.633 | 0.569 |
| 1.966 | cat | 171 | 13G | 0.584 | 0.509 |
| weighted impotant | | | | **0.596** | **0.558** |

Table 5: The most important models on the reverse-calibrated dataset for the metamodel and their characteristics

| Impotant scr | algo | test | val | Class 0 | Class 1 |
|---|---|---|---|---|---|
| 37.392 | lgb | 12G | 171 | 0.775 | 0.642 |
| 36.928 | cat | 12G | 171 | 0.797 | 0.666 |
| 18.042 | cat | 165 | 12G | 0.772 | 0.677 |
| 12.090 | lgb | 165 | 12G | 0.720 | 0.622 |
| 11.602 | xgb | 14G | 165 | 0.753 | 0.560 |
| 1.460 | cat | 14G | 165 | 0.707 | 0.511 |
| weighted impotant | | | | **0.772** | **0.643** |

The results of the evaluation show an increase in the quality of class prediction after reverse-calibration of initial data from 0.558 to 0.643 (Table 4, Table 5). Average weighted metric F1 score of reservoir prediction quality at the level of 0.643 is acceptable for the given stage and nature of the field study (dominance of data from horizontal wells). This F1 score metric value does not correlate with the absolute value of metamodel prediction quality and is used for relative comparison of algorithm performance on data sets. In practice, the quality of the metamodel predictions exceeds the quality of the predictions of the individual models and their weighted averages.

As a result of model ensemble and calibration of metamodel probabilities on isotonic regression - predictions are made. A three-dimensional cube of the calibrated probabilities of the space under study belonging to the "manifold" class is obtained.

To analyze the three-dimensional probability cube, we calculated the map of net pay thicknesses over the entire Achimov complex of sediments (Fig. 1). The effective thicknesses were calculated as follows:

$$h_{ij} = \sum_{k=1}^{K} 1_{(P_{ijk} \geq P_T)} s,$$

where $s$ is the vertical resolution of the approximation grid, $P_{ijk}$ is the predicted probability that the voxel belongs to the collector class, $P_T$ is the threshold probability that the voxel belongs to the collector class, and $1_{(\cdot)}$ is the indicator function.

The vertical resolution of the approximation grid $s$ of 5 meters per voxel was used to build the map, and the probability threshold $P_T$ was set to 0.5.

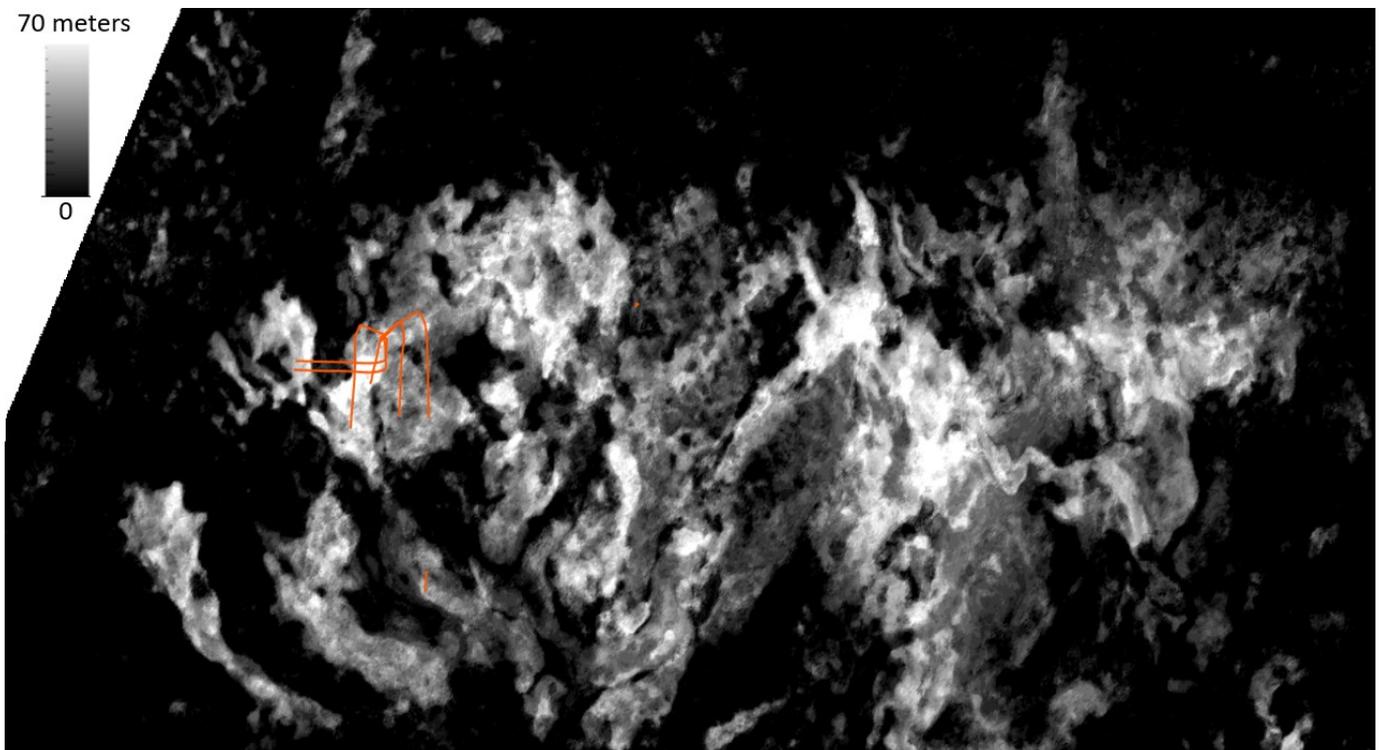

Figure 2: Map of predicted reservoir thicknesses of the Achimov sedimentary complex, derived from a reversed-calibrated dataset. Orange curves - well trajectories

## Conclusions

The map of predicted reservoir thicknesses traces the general patterns of the structure architecture of the Achimov sedimentation complex: landslide bodies characteristic of this sedimentation complex, turbidite flows of different density and areas of sediment discharge are contrastingly highlighted. This prediction depends on the results of lithological interpretation of the logging data; thus, if the reservoir cutoff changes according to petrophysical properties, the probabilistic representation of the geological object will change.

The reverse-calibration method proposed in this paper uses uncertainties in the geological and geophysical data as a hyperparameter for global tuning of the technological stack from machine learning algorithms, through modification of the training dataset. The boundaries of the reverse-calibration response search are set based on an a priori assumption about the limits of our knowledge of the natural object and the possible instrumental errors in obtaining this knowledge. In this work, we used the search for position displacement of the seismic wavefield relative to the well trajectories and the scale of data approximation to the grid. It is shown that for this variant, the method increases the values of weighted metric F1 of the metamodel measure from 0.558 to 0.643. At the same time, the reverse-calibration method can vary other parameters of training sample acquisition by reverse-engineering the data. For example, the cutoff by petrophysical properties of reservoir extraction can be used as another hyperparameter of algorithm stack tuning, through modification of the training sample.

It should be noted that the low absolute values of quality metrics of individual classification models obtained as a result of the study may indicate the complexity of the object for generalization of borehole and seismic data at the current stage of the field study. Thus, for such a complex object of study as the Achimov complex of deposits, the obtained three-dimensional probabilistic representation of the reservoir can be used as an auxiliary tool, i.e., a tool of primary formalization of the accumulated knowledge about the object of study on the basis of the available information. The field studied in the work is at the early stage of study, so its further development will require drilling of exploration wells in the zones of geological interest. Selection of these zones can be carried out with the help of the proposed approach. After drilling the wells and obtaining new data, the model can be updated and used for more accurate assessment of the geological object, which in turn improves decision making on further exploration and provides a continuous cycle of independent assessments based on actual data with positive feedback from any results of drilling each new well.

The approach proposed in the work based on the technological stack of machine learning algorithms allows to formalize the primary generalization of data about the object of study, considering the existing uncertainties, using the actual material in the form of well survey results and geospatial data. The approach may be one of the tools for generalization of available geological and geophysical information at the existing level of territory study. With the help of this tool it is possible to create three-dimensional probabilistic representation of the geological object, to improve this representation after receiving new data, to create a continuous cycle of formalized evaluations. The results of the forecast can be used as an expert-independent initial hypothesis for the subsequent work of experts, algorithms and decision-making.

# List of references

1. BoostARoota GitHub repository. https://github.com/chasedehan/BoostARoota.
2. Prokhorenkova G., Gusev A., Vorobev A., Dorogush, Gulin A. CatBoost: unbiased boosting with categorical features. In Bengio S., Wallach H., Larochelle H., Grauman K., Cesa-Bianchi N., Garnett R. editors. Proceedings of the 31st International Conference on Advances in Neural Information Processing Systems (NeurIPS'18). Curran Associates, 2018.
3. Ivlev D. Reservoir Prediction by Machine Learning Methods on The Well Data and Seismic Attributes for Complex Coastal Conditions arXiv preprint arXiv:2301.03216v1 2023.
4. Bergstra James S., et al. Algorithms for hyper-parameter optimization. Advances in Neural Information Processing Systems. 2011.
5. Bergstra James, Daniel Yamins, David Cox. Making a science of model search: Hyperparameter optimization in hundreds of dimensions for vision architectures. Proceedings of The 30th International Conference on Machine Learning, 2013.
6. Ke Q. Meng, T. Finley, T. Wang, W. Chen, W. Ma, Q. Ye, T.-Y. Liu. Lightgbm: A highly efficient gradient boosting decision tree. In I. Guyon, U. von Luxburg, S. Bengio, H. Wallach, R. Fergus, S. Vishwanathan, and R. Garnett editors. Proceedings of the 30th International Conference on Advances in Neural Information Processing Systems (NeurIPS'17). Curran Associates, 2017.
7. Chen and C. Guestrin. Xgboost: A scalable tree boosting system. In B. Krishnapuram, M. Shah, A. Smola, C. Aggarwal, D. Shen, and R. Rastogi editors. Proceedings of the 22nd ACM SIGKDD International Conference on Knowledge Discovery and Data Mining (KDD'16), ACM Press, 2016, pp. 785–794.
8. Hollmann N., Müller S., Eggensperger K., Hutter F. TabPFN: A Transformer That Solves Small Tabular Classification Problems in a Second. arXiv preprint arXiv: 2207.01848v4, 2022.
9. Scott Lundberg, Su-In Lee. A Unified Approach to Interpreting Model Predictions arXiv arXiv:1705.07874v2, 2017.